\RequirePackage{fix-cm}
\documentclass[smallextended]{svjour3}       
\smartqed  
\usepackage{amsfonts,amsmath,amssymb}
\usepackage{indentfirst, setspace}
\usepackage{url,float}
\usepackage{color}
\usepackage[a4paper,ignoreall]{geometry}
\usepackage{longtable}
\usepackage{multicol}
\usepackage{stfloats}
\usepackage{enumerate}
\usepackage{cite}
\usepackage{multirow}
\usepackage{amssymb}
\usepackage{booktabs}
\usepackage[colorlinks]{hyperref}
\usepackage{textcomp}
\usepackage{algorithm}
\usepackage{algorithmic}
\usepackage[justification=centering]{caption}
\usepackage{mathtools}
\usepackage{mhchem}
\usepackage{tensor}
\def\qu#1 {\fbox {\footnote {\ }}\ \footnotetext { From Qu: {\color{red}#1}}}
\def\hqu#1 {}

\usepackage[misc]{ifsym}

\newtheorem{lem}{Lemma}

\newtheorem{defn}{Definition}

\newtheorem{con}{Conjecture}

\newcommand{\F}{\mathbb {F}}

\begin{document}

\title{On two conjectures about the intersection distribution}



\author{Yubo Li \textsuperscript{1} \and Kangquan Li \textsuperscript{1} \and Longjiang Qu \textsuperscript{1, 2} }


\institute{\Letter~ Longjiang Qu\\
\email{leeub\_0425@hotmail.com \and likangquan11@nudt.edu.cn \and ljqu\_happy@hotmail.com}\\
$\prescript{1}{}{~\text{College of Liberal Arts and Sciences, National University of Defense Technology, Changsha, 410073, China.}}$\\
$\prescript{2}{}{~\text{State Key Laboratory of Cryptology, Beijing, 100878, China.}}$}


\date{Received: date / Accepted: date}

\maketitle

\begin{abstract}
Recently, S. Li and A. Pott\cite{LP} proposed a new concept of intersection distribution concerning the interaction between the graph $\{(x,f(x))~|~x\in\F_{q}\}$ of $f$ and the lines in the classical affine plane $AG(2,q)$. 
Later, G. Kyureghyan, et al.\cite{KLP} proceeded to consider the next simplest case and derive the intersection distribution for all degree three polynomials over $\F_{q}$ with $q$ both odd and even.
They also proposed several conjectures in \cite{KLP}.

In this paper, we completely solve two conjectures in \cite{KLP}. 
Namely, we prove two classes of power functions having intersection distribution: $v_{0}(f)=\frac{q(q-1)}{3},~v_{1}(f)=\frac{q(q+1)}{2},~v_{2}(f)=0,~v_{3}(f)=\frac{q(q-1)}{6}$.
We mainly make use of the multivariate method and QM-equivalence on $2$-to-$1$ mappings. 
The key point of our proof is to consider the number of the solutions of some low-degree equations.

\end{abstract}

\section{Introduction}
Let $\F_{q}=\F_{p^m}$ be a finite field with characteristic $p$ and $f$ a polynomial over $\F_{q}$.
S. Li and A. Pott\cite{LP} proposed a new concept of intersection distribution as follows.
\begin{defn}
	\textup{(Intersection distribution)}
	For $0\le i\le q$, define $$v_{i}(f)=|\{(b, c)\in\F_{q}^{2}~|~f(x)-bx-c=0 \textit{~has~} i \textit{~solutions in~} \F_{q}\}|.$$
	The sequence $(v_{i}(f))^{q}_{i=0}$ is the intersection distribution of $f$. 
	The integer $v_{0}(f)$ is the non-hitting index of $f$.
\end{defn}

The intersection distribution of a polynomial $f$ originates from an elementary problem concerning the interaction between the graph $\{(x,f(x))~|~x\in\F_{q}\}$ of $f$ and the lines in the classical affine plane $AG(2,q)$. 
More precisely, for $0\le i\le q$, one may ask about the number of affine lines intersecting the graph of $f$ in exactly $i$ points.
When $q$ is even, the long-standing open problem of classifying $o$-polynomials can be rephrased in a simple way, namely, classifying all polynomials which have the same intersection distribution as $x^2$.
Meanwhile, S. Li and A. Pott\cite{LP} also identified a particularly interesting quantity named non-hitting index. 
The non-hitting index $v_{0}(f)$ measures the distance from $f$ to linear functions, and to the so called $o$-polynomial (when $q$ is even) or to $x^2$ (when $q$ is odd)\cite[Result 1.7]{LP}.

Recently, G. Kyureghyan, et al.\cite{KLP} proceeded to consider the next simplest case and derive the intersection distribution for all degree three polynomials over $\F_{q}$ with $q$ both odd and even. 
Moreover, they initiated to classify all monomials having the same intersection distribution as $x^3$, where some characterizations of such monomials were obtained and some conjectures were proposed.
\begin{con}
\label{Conjecture}
\cite[Conjecture 3.2(1)]{KLP}
The following two families of monomials $f(x)=x^d$ over $\F_{q}=\F_{3^m}$,
	
\begin{enumerate}[(i)]
\item $d=3^{\frac{m+1}{2}}+2$ and $d^{-1}$, $m$ odd;
\item $d=2\cdot3^{m-1}+1$ and $d^{-1}$, $m$ odd;
\end{enumerate} 
have intersection distribution
\begin{equation}
\label{distribution}
v_{0}(f)=\frac{q(q-1)}{3},~v_{1}(f)=\frac{q(q+1)}{2},~v_{2}(f)=0,~v_{3}(f)=\frac{q(q-1)}{6}.
\end{equation}
\end{con}

In this paper, we completely solve Conjecture \ref{Conjecture}.
According to Theorem 3.8 of \cite{KLP} (see also Lemma \ref{2-1 trans}), the key point of proving Conjecture \ref{Conjecture} is to establish that some polynomial $g_{d}(x)=\frac{x^d-1}{x-1}$ is $2$-to-$1$ over $\F_{q}\setminus\{1\}$, where $q=3^m$.
One can refer to \cite{LQM,MQ} for more details about $2$-to-$1$ mappings.
In \cite{LQM}, the authors introduced  a quasi-multiplicatively (QM) equivalence between two $2$-to-$1$ mappings as follows. Two polynomials $h(x)$ and $g(x)$ in $\F_{q}[x]$ are said to be QM equivalent if there exists an integer $1\le d\le q-1$ with $\gcd(d,q-1)=1$ and $h(x)=ag(bx^d)$ for some nonzero elements $a,b\in\F_q$.
In our proof, we do not directly prove the $2$-to-$1$ property of $g_{d}(x)=\frac{x^d-1}{x-1}$, but consider that of another polynomial that is QM equivalent to $g_{d}(x)$.
Moreover, in the process of proof, we mainly use the multivariate method introduced by H. Dobbertin \cite{Dobbertin} to verify the $2$-to-$1$ property.
The multivariate method has been applied to prove APN functions and permutation polynomials before, and one may refer to \cite{D-Welch,D-Niho,Dobbertin-5,LQCL,WZZ} for more details.
Finally, the proof can be transformed into discussing the number of the solutions of some low-degree equations.

The rest of the paper is organized as follows.
In Section 2, we introduce some relevant results that will be frequently used.
In Section 3, we give the complete proof of Conjecture \ref{Conjecture}.
Finally, Section 4 is the conclusion.

\section{Preliminaries}

In this section, we give necessary definitions and results which will be frequently used in this paper.
Throughout this paper, denote $\F_{q}^{*}$ the set of nonzero elements in $\F_{q}$.
\subsection{Multiplicity distribution and relevant results}
Firstly, to facilitate the computation of the intersection distribution, we introduce the following definition which was proposed in \cite[Definition 1.1(2)]{LP}.
\begin{defn}
\textup{(Multiplicity distribution)}
Let $f$ be a polynomial over $\F_{q}$. 
For $b\in\F_{q}$ and $0\le i\le q$, define $$M_{i}(f,b)=|\{c\in\F_{q}~|~f(x)-bx-c=0 \textit{~has~} i \textit{~solutions in~} \F_{q}\}|.$$
The sequence $(M_{i}(f,b))^{q}_{i=0}$ is the multiplicity distribution of $f$ at $b$. 
The multi-set of sequences\\ $\{(M_{i}(f,b))^{q}_{i=0}~|~b\in\F_{q}\}$ is the multiplicity distribution of $f$.
\end{defn}

By the definition, for $0\le i\le q$, there are exactly $M_{i}(f,b)$ lines among the parallel class of $q$ affine lines $\{y=bx+c~|~c\in\F_{q}\}$, which intersect the graph of $f$ in $i$ points.
Based on this definition, we show a useful result.

\begin{lem}
\cite[Remark 1.4(3)]{KLP}
\label{PP}
Let $f$ be a permutation polynomial and $f^{-1}$ be its inverse. 
Clearly, $M_{1}(f,0)=M_{1}(f^{-1},0)=q$.
Moreover, note that for $b\in\F_{q}^{*}$, the two equations $f(x)-bx-c=0$ and $\displaystyle f^{-1}(x)-\frac{1}{b}x+\frac{c}{b}=0$ have the same number of solutions. 
Hence, $f$ and $f^{-1}$ have the same multiplicity distribution and therefore, the same intersection distribution.
\end{lem}

Furthermore, G. Kyureghyan, et al.\cite{KLP} gave the following necessary and sufficient condition characterizing monomials over $\F_{q}$ satisfying (\ref{distribution}), when $q$ is a power of 3.
\begin{lem}
\cite[Theorem 3.8]{KLP}
\label{2-1 trans}
Let $f(x)$ be over $\F_{q}=\F_{3^m}$. Then $f$ satisfies (\ref{distribution}) if and only if for each $y\in\F_{q}$, the function $\frac{f(x+y)-f(y)}{x}\Big\arrowvert_{\F_{q}^{*}}$ is $2$-to-$1$. In particular, $f(x) = x^d$ satisfies (\ref{distribution}) if and only if the following holds:
\begin{enumerate}[(i)]
\item $\gcd(d-1,q-1)=2$;
\item $g_{d}\big\arrowvert_{\F_{q}\setminus\{1\}}$ is $2$-to-$1$, where $g_{d}(x)=\displaystyle\frac{x^d-1}{x-1}$.
\end{enumerate}
\end{lem}

\subsection{Solutions of low-degree equations}
In this subsection, we mainly introduce a known lemma about the solutions of the equation with degree three, which will be used in the proofs of our results.
When $\varphi$ is reducible over $\F_{q}$, it is abbreviated as $\varphi = (1, 1, 1)$ if it can be factorized as three linear factors, and $\varphi = (1, 2)$ if it can be factorized as a product of a linear factor and an irreducible quadratic factor. 
When $\varphi$ is irreducible over $\F_{q}$, it is abbreviated as $\varphi = (3)$.
For two positive integers $m$ and $n$ with $m| n$, we use ${\rm{Tr}}_{m}^{n}(\cdot)$ to denote the trace function from $\F_{p^n}$ to $\F_{p^m}$, i.e., $${\rm{Tr}}_{m}^{n}(x):=x+x^{p^m}+x^{p^{2m}}+\cdots+x^{p^{(\frac{n}{m}-1)m}}.$$
Particularly, when $m=1$, it is the absolutely trace function.
\begin{lem}
\label{solution}
\cite{Williams}
The factorization of $\varphi(x)=x^3+ax+b~(a,b\in\F_{3^n})$ over $\F_{3^n}$ are characterized as follows:
\begin{enumerate}[(i)]
\item $\varphi = (1, 1, 1)$ if and only if $-a$ is a square in $\F_{3^n}$, say $-a = c^2$, and ${\rm{Tr}}_{1}^{n}(\frac{b}{c^3})=0$;
\item $\varphi = (1, 2)$ if and only if $-a$ is not a square in $\F_{3^n}$;
\item $\varphi = (3)$ if and only if $-a$ is a square in $\F_{3^n}$, say $-a = c^2$, and ${\rm{Tr}}_{1}^{n}(\frac{b}{c^3})\neq0$.
\end{enumerate}
\end{lem}

\begin{lem}
\label{gcd=2}
\cite[Lemma 9]{EFPST}
Let $p,k,m$ be integers greater than or equal to 1 (we take $k\le m$, though the result can be shown in general).
Then $$\gcd(p^k+1,p^m-1)=2, \textit{~if~} \frac{m}{\gcd(m,k)} \textit{~is odd}.$$
\end{lem}
Consequently, if $m$ is odd, then $\gcd(p^k+1,p^m-1)=2$ holds naturally.

\subsection{Resultant of polynomials}
In this subsection, we recall some basic facts about the resultant of two polynomials. 
Given two non-zero polynomials of degrees $n$ and $m$ respectively
$$u(x)=a_{m}x^m+a_{m-1}x^{m-1}+\cdots+a_{0},~~ v(x)=b_{n}x^n+b_{n-1}x^{n-1}+\cdots+b_{0},$$
with $a_{m}\neq0,b_{n\neq0}$, and coefficients in a field or in an integral domain $\mathbb{R}$, their resultant Res$(u,v)\in\mathbb{R}$ is the determinant of the following matrix:
\begin{displaymath}
	\left( \begin{array}{cccccccccc}
		a_{m} & a_{m-1} & \ldots & \ldots & \ldots & \ldots & a_{0} & 0 & 0 & 0 \\
		0 & a_{m} & a_{m-1} & \ldots & \ldots & \ldots & \ldots & a_{0} & 0 & 0 \\
		& & \ddots & \ddots & & & & & \ddots & \\
		0 & 0 & 0 & a_{m} & a_{m-1} & \ldots & \ldots & \ldots & \ldots & a_{0} \\
		b_{n} & b_{n-1} & \ldots & \ldots & b_{0} & 0 & 0 & \ldots & \ldots & 0 \\
		& \ddots & \ddots & & & & \ddots & & & \\
		0 & 0 & \ldots & \ldots & 0 & b_{n} & b_{n-1} & \ldots & \ldots & b_{0}
	\end{array} \right).
\end{displaymath}

For a field $\mathbb{K}$ and two polynomials $F(x,y),G(x,y)\in\mathbb{K}[x,y]$, we use Res$(F,G,y)$ to denote the resultant of $F$ and $G$ with respect to $y$. 
It is the resultant of $F$ and $G$ when considered as polynomials in the single variable $y$. 
In this case, Res$(F,G,y)\in\mathbb{K}[x]$ belongs in the ideal generated by $F$ and $G$, and thus any $a,b$ satisfying $F(a,b)=0$ and $G(a,b)=0$ is such that Res$(F,G,y)(a)=0$ (see \cite{LN} for more details).

\section{The Proof of Conjecture \ref{Conjecture}}
In this section, we present the complete proof of Conjecture \ref{Conjecture}.
\subsection{$d=3^{\frac{m+1}{2}}+2$ and $d^{-1}$, $m$ odd}

\textit{Proof of Conjecture \ref{Conjecture}(i).}
We divide the proof into two parts: $d$ and $d^{-1}$.

\textbf{Part 1.}
According to Lemma \ref{2-1 trans}, for $f(x)=x^d$, we need to show that $\gcd(d-1,q-1)=2$ and $g_{d}\big\arrowvert_{\F_{q}\setminus\{1\}}$ is $2$-to-$1$, where $q=3^m$ and $g_{d}(x)=\displaystyle\frac{x^{d}-1}{x-1}$.

Firstly, it is straightforward by Lemma \ref{gcd=2} that $\gcd(d-1,q-1)=(3^{\frac{m+1}{2}}+1,3^m-1)=2$, since $m$ is odd.

Secondly, we will establish that $g_{d}(x)=\displaystyle\frac{x^{3^{\frac{m+1}{2}}+2}-1}{x-1}$ is $2$-to-$1$ over $\F_{q}\setminus\{1\}$.
Assume that $$h_{1}(x)=g_{d}\Big(x^{3^{\frac{m-1}{2}}}\Big)=\frac{x^{2\cdot3^{\frac{m-1}{2}}+1}-1}{x^{3^{\frac{m-1}{2}}}-1}.$$
With the fact that $\gcd(3^{\frac{m-1}{2}},q-1)=\gcd(3^{\frac{m-1}{2}},3^m-1)=1$, we know that $x^{3^{\frac{m-1}{2}}}$ is a permutation over $\F_{q}$, then $h_{1}(x)$ is 2-to-1 over $\F_{q}\setminus\{1\}$ if and only if so is $g_{d}(x)$.

Let $y=x^{3^{\frac{m-1}{2}}}$ and $b=a^{3^{\frac{m-1}{2}}}$ for any $a\in\F_{3^m}$.
According to the definition, $h_{1}(x)$ is 2-to-1 if and only if $h_{1}(x+a)-h_{1}(a)=0$ has exactly two solutions $x=0$ and $x=x_{0}\in\F_{3^m}^{*}$ for any $a\in\F_{3^m}\setminus\{1\}$, where $x+a\neq1$.

Then $h_{1}(x)=\frac{y^2x-1}{y-1}$ and $h_{1}(x+a)-h_{1}(a)=\frac{(y+b)^2(x+a)-1}{y+b-1}-\frac{b^2a-1}{b-1}$.
Since $y+b-1\neq0$ and $b-1\neq0$, it suffices to show that for any $a\in\F_{3^m}\setminus\{1\}$,
$$h_{1}(x+a)-h_{1}(a)=0,$$
after computing and simplifying, i.e.,
\begin{equation}
\label{yuanshi-1}
(b-1)(y^2x+2bxy+b^2x+ay^2+2aby)-(b^2a-1)y=0
\end{equation}
has exactly two solutions in $\F_{3^m}\setminus\{1\}$.
Meanwhile, $x=0$ is always a solution of Eq. (\ref{yuanshi-1}), thus, we mainly consider the solution $x\neq0$ in the following.

Raising Eq. (\ref{yuanshi-1}) into its $3^{\frac{m+1}{2}}$-th power and simplifying it by $y^{3^{\frac{m+1}{2}}}=x,x^{3^{\frac{m+1}{2}}}=y^3$ and $b^{3^{\frac{m+1}{2}}}=a,a^{3^{\frac{m+1}{2}}}=b^3$, we have
\begin{equation}
\label{raising-1}
(a-1)(x^2y^3+2ay^3x+a^2y^3+b^3x^2+2b^3ax)-(a^2b^3-1)x=0
\end{equation}

When $a=0$, then $b=0$. 
From Eq. (\ref{yuanshi-1}), we obtain that $-y^2x+y=0$, i.e., $xy=1$.
Then by Eq. (\ref{raising-1}), $y^3x=1$.
We get $y^2=1$ and $y=-1$, which is the unique nonzero solution of Eq. (\ref{yuanshi-1}).

When $a=2$, then $b=2$.
Eq. (\ref{yuanshi-1}) is equivalent to $$y^2x+xy+x+2y^2+y=0.$$
That is, $$x(y-1)^2=y(y-1).$$
If $y=1$, the equation holds naturally.
If $y\neq1$, $xy=x+y$.
Furthermore, we have $y^3x=y^3+x$, which leads to $y^2=1$.
Then $y=-1$, leading $y+b=1$, which is a conflict.
Thus, $y=1$ is the unique nonzero solution of Eq. (\ref{yuanshi-1}) at this moment.

In the rest of this proof, we assume that $a\in\F_{3^m}\setminus\F_{3}$.
Let
$$F(x,y)=(b-1)(y^2x+2bxy+b^2x+ay^2+2aby)-(b^2a-1)y$$
and $$G(x,y)=(a-1)(x^2y^3+2ay^3x+a^2y^3+b^3x^2+2b^3ax)-(a^2b^3-1)x.$$
After computing by MAGMA, we have
$$\textup{Res}(F,G,y)=x(x+a)^2(x+a+2)\Big((b+2)^3(a^2b^3+2)^2x-(ab^2+ab+1)^3(a^2b^3+ab^3+1)\Big).$$
Therefore, from Eq. (\ref{yuanshi-1}) and Eq. (\ref{raising-1}), $x=-a$ or $x=-a-2$ or $(b+2)^3(a^2b^3+2)^2x-(ab^2+ab+1)^3(a^2b^3+ab^3+1)=0$.
In the following, we claim $x\neq-a$ and $x\neq-a-2$.

If $x=-a$, then $y=-b$.
From Eq. (\ref{yuanshi-1}), we obtain that $ab=1$.
With $b^3a=1$, then $b^2=1$ holds, which contradicts with the assumption $a\in\F_{3^m}\setminus\F_{3}$.

If $x=-a-2$, then $x+a=1$, which contradicts with the original assumption $x+a\neq1$.

Hence, it
suffices to show $(b+2)^3(a^2b^3+2)^2\neq0$ for $a\in\F_{3^m}\setminus\F_{3}$.
Firstly, $b+2\neq0$, which is trivial.
Secondly, assume that there exists some $a$ such that $$a^2b^3=1.$$
Meanwhile, $b^6a^3=1$ also holds, which leads to $a=1$ and contradicts.
Thus $x=\frac{(ab^2+ab+1)^3(a^2b^3+ab^3+1)}{(b+2)^3(a^2b^3+2)^2}$. 

In the sequel, we will show that $(ab^2+ab+1)^3(a^2b^3+ab^3+1)\neq0$ and $x\neq-a+1$.
Since
$(a^2b^3+ab^3+1)=(ab^2+ab+1)^{3^{\frac{m+1}{2}}}$, we only need to consider when $ab^2+ab+1$ is not equal to $0$.
Suppose the contrary, let $ab^2+ab+1=0$.
Then we have $a=\frac{2}{b^2+b}$.
Plugging $a=\frac{2}{b^2+b}$ into $a^2b^3+ab^3+1=0$, we obtain $b^3=1$, which leads $b=1$ and a contradiction.
Next, assume $x=\frac{(ab^2+ab+1)^3(a^2b^3+ab^3+1)}{(b+2)^3(a^2b^3+2)^2}=-a+1$.
Similarly, we can also deduce that $b=1$, which is a conflict.

Therefore, $x=\frac{(ab^2+ab+1)^3(a^2b^3+ab^3+1)}{(b+2)^3(a^2b^3+2)^2}$ is the unique nonzero solution of Eq. (\ref{yuanshi-1}), and we are done.

\textbf{Part 2.} 
As for $d^{-1}$, we first show that $x^d=x^{3^{\frac{m+1}{2}}+2}$ is a permutation over $\F_{q}$.
By Euclidean algorithm $$3^m-1=(3^{\frac{m+1}{2}}+2)(3^{\frac{m-1}{2}}-1)+3^{\frac{m-1}{2}}+1,$$
and $$3^{\frac{m+1}{2}}+2=(3^{\frac{m-1}{2}}+1)\cdot3-1,$$
we conclude that $\gcd(d,q-1)=\gcd(3^{\frac{m+1}{2}}+2,3^m-1)=1$.
Furthermore, $x^{d^{-1}}$ is the inverse of $x^d$. 
Then by Lemma \ref{PP}, we know that $x^{d^{-1}}$ has the same intersection distribution with $x^d$.

Thus the proof is completed.\hfill$\square$

\subsection{$d=2\cdot3^{m-1}+1$ and $d^{-1}$, $m$ odd}

\textit{Proof of Conjecture \ref{Conjecture}(ii).}
We also divide the proof into two parts: $d$ and $d^{-1}$.

\textbf{Part 1.}
According to Lemma \ref{2-1 trans}, for $f(x)=x^d$, we need to show that $\gcd(d-1,q-1)=2$ and $g_{d}\big\arrowvert_{\F_{q}\setminus\{1\}}$ is $2$-to-$1$, where $q=3^m$ and $g_{d}(x)=\displaystyle\frac{x^{d}-1}{x-1}$.

Firstly, $\gcd(d-1,q-1)=\gcd(2\cdot3^{m-1},3^m-1)=\gcd(2\cdot3^{m},3^m-1)=\gcd(2,3^m-1)=2$.

Secondly, we will establish that $g_{d}(x)=\displaystyle\frac{x^{2\cdot3^{m-1}+1}-1}{x-1}$ is $2$-to-$1$ over $\F_{q}\setminus\{1\}$.
Assume that $$h_{2}(x)=g_{d}(x^{3})=\frac{x^{2\cdot3^{m}+3}-1}{x^{3}-1}=\frac{x^{5}-1}{x^{3}-1}=\frac{x^4+x^3+x^2+x+1}{x^2+x+1}.$$
Obviously, $x^{3}$ is a permutation over $\F_{q}$, then $h_{2}(x)$ is $2$-to-$1$ over $\F_{q}\setminus\{1\}$ if and only if so is $g_{d}(x)$.

According to the definition, $h_{2}(x)$ is $2$-to-$1$ if and only if $h_{2}(x+a)-h_{2}(a)=0$ has exactly two solutions $x=0$ and $x=x_{0}\in\F_{3^m}^{*}$ for any $a\in\F_{3^m}\setminus\{1\}$, where $x+a\neq1$.
In this sense, it suffices to show that, for any $a\in\F_{3^m}\setminus\{1\}$,
$$h_{2}(x+a)-h_{2}(a)=\frac{(x+a)^4+(x+a)^3+(x+a)^2+(x+a)+1}{(x+a)^2+(x+a)+1}-\frac{a^4+a^3+a^2+a+1}{a^2+a+1}=0,$$
after computing and simplifying, i.e.,
\begin{equation}
	\label{yuanshi-2}
	(a^2+a+1)x^4+(a^3+2a^2+2a+1)x^3+(2a^4+2a^3)x^2+(2a^5+a^4)x=0
\end{equation}
has exactly two solutions in $\F_{3^m}\setminus\{1\}$.
Meanwhile, $x=0$ is always a solution of Eq. (\ref{yuanshi-2}).
Thus, \begin{equation}
	\label{3ci}
	x^3+(a+1)x^2+\frac{2a^4+2a^3}{a^2+a+1}x+\frac{2a^5+a^4}{a^2+a+1}=0
\end{equation}
has exactly one solution in $\F_{3^m}\setminus\{0,1\}$ if $h_{2}(x)$ is $2$-to-$1$.

If $a=0$, then Eq. (\ref{3ci}) becomes $x^3+x^2=0$, which leads $x=-1$. Thus, $x=-1$ is the unique nonzero solution.
If $a=2$, then Eq. (\ref{3ci}) becomes $x^3-1=0$, which leads $x=1$. Thus, $x=1$ is the unique nonzero solution.

Hence, in the rest of this proof, we assume that $a\in\F_{3^m}\setminus\F_{3}$.
At this time, $a+1\neq0$ and $\frac{2a^5+a^4}{a^2+a+1}\neq0$.
Let $$H(x)=x^3+(a+1)x^2+\frac{2a^4+2a^3}{a^2+a+1}x+\frac{2a^5+a^4}{a^2+a+1}.$$
Then $$\widetilde{H}(x)=x^3H\Big(\frac{1}{x}+\frac{2a^3}{a^2+a+1}\Big)=\frac{a^5+a^4}{(a^2+a+1)^3}x^3+(a+1)x+1.$$
Assume $$M(x)=\frac{(a^2+a+1)^3}{a^5+a^4}\widetilde{H}(x)=x^3+\frac{(a^2+a+1)^3}{a^4}x+\frac{(a^2+a+1)^3}{a^5+a^4}.$$
At this moment, $$-\frac{(a^2+a+1)^3}{a^4}=-\frac{(a^2+a+1)^2(a-1)^2}{a^4}.$$
However, $(-1)^{\frac{3^m-1}{2}}\neq1$, since $m$ is odd. 
Then $-1$ is a non-square in $\F_{3^m}$, and so is the $-\frac{(a^2+a+1)^3}{a^4}$.
According to Lemma \ref{solution}, we conclude that $M(x)$ has a unique solution in $\F_{3^m}$, and so is $H(x)$.
Hence, $h_{2}(x)$ is $2$-to-$1$ over $\F_{q}\setminus\{1\}$.

\textbf{Part 2.} 
As for $d^{-1}$, we first show that $x^d=x^{2\cdot3^{m-1}+1}$ is a permutation over $\F_{q}$.
By Euclidean algorithm, $$\gcd(d,q-1)=\gcd(2\cdot3^{m-1}+1,3^m-1)=\gcd(2\cdot3^{m-1}+1,3^{m-1}-2)=\gcd(3^{m-1}-2,5).$$
Since $m$ is odd, then the last digit of $3^{m-1}$ is $1$ or $9$.
It is straightforward that $\gcd(3^{m-1}-2,5)=1$, and so is the $\gcd(d,q-1)=1$.
Furthermore, $x^{d^{-1}}$ is the inverse of $x^d$. 
Then by Lemma \ref{PP}, we know that $x^{d^{-1}}$ has the same intersection distribution with $x^d$.

Thus the proof is completed.\hfill$\square$

\section{Conclusion}
In this paper, we completely solve Conjecture \ref{Conjecture} that was proposed by G. Kyureghyan, et al. \cite{KLP}, by using the multivariate method introduced by H. Dobbertin \cite{Dobbertin} and QM-equivalence on $2$-to-$1$ mappings. 
The key point of our proof is to consider the number of the solutions of some low-degree equations.
Future potential investigations may go in the direction of finding theoretical explanations for more non-hitting indices and searching more classes of functions having intersection distribution (\ref{distribution}).
Furthermore, we will try to prove \cite[Conjecture 3.2(2)]{KLP}.

\section*{Acknowledgments}
The authors would like to thank A. Prof. Yue Zhou for providing this research topic and giving many helpful suggestions which improved the clarity and the quality of the paper a lot.


\begin{thebibliography}{99}

\bibitem{D-Welch}
H. Dobbertin. Almost perfect nonlinear power functions on GF($2^n$): the Welch case. IEEE Transactions on Information Theory, 45 (4), pp. 1271-1275, 1999.

\bibitem{D-Niho}
H. Dobbertin. Almost perfect nonlinear power functions on GF($2^n$): the Niho case. Information and Computation, 151 (1-2), pp. 57-72, 1999.
	
\bibitem{Dobbertin}
H. Dobbertin. Uniformly representable permutation polynomials. Sequence and their Applications-SETA 2001, Springer, 2, pp. 1-22, 2002.

\bibitem{Dobbertin-5}
H. Dobbertin. Almost perfect nonlinear power functions on GF($2^n$): A New Case for $n$ Divisible by $5$. In: Finite Fields and Applications, Augsburg, Springer, Berlin, pp. 113-121, 2001. DOI: 10.1007/978-3-642-56755-1\_11.
	
\bibitem{EFPST}
P. Ellingsen, P. Felke, C. Riera, P. St$\breve{\textup{a}}$nic$\breve{\textup{a}}$, A. Tkachenko. $C$-differentials, multiplicative uniformity and (almost) perfect $c$-nonlinearity. IEEE Transactions on Information Theory, 2020. DOI: 10.1109/TIT.2020.2971988.
	
\bibitem{KLP}	
G. Kyureghyan, S. Li and A. Pott. On the intersection distribution of degree three polynomials and related topics. ArXiv:2003.10040v1.

\bibitem{LQM}
K. Li, S. Mesnager and L. Qu. Further study of 2-to-1 mappings over $\F_{2^n}$. ArXiv:1910.06654v1.

\bibitem{LP}
S. Li and A. Pott. Intersection distribution, non-hitting index and Kakeya sets in affine planes. Finite Fields and Their Applications, 2020. DOI: 10.1016/j.ffa.2020.101691.

\bibitem{LQCL}
K. Li, L. Qu, X. Chen and C. Li. Permutation polynomials of the form $cx+{\rm{Tr}}_{q^l/q}(x^a)$ and permutation trinomials over finite fields with even characteristic. Cryptography and Communications, 10, pp. 531-554, 2018.
	
\bibitem{LN}
R. Lidl and H. Niederreiter. Finite Fields, 2nd ed. Cambridge University Press, Cambridge, 1997.
		
\bibitem{MQ}
S. Mesnager and L. Qu. On two-to-one mappings over finite fields. IEEE Transactions on Information Theory, 65 (12), pp. 7884-7895, 2019.
	
\bibitem{WZZ}
Y. Wang, W. Zhang and Z. Zha. Six new classes of permutation trinomials over $\mathbb{F}_{2^{n}}$. SIAM Journal on Discrete Mathematics, 32 (3), pp. 1946-1961, 2018.
	
\bibitem{Williams}
K. Williams. Note on cubics over GF($2^n$) and GF($3^n$). Journal of Number Theory, 7 (4), pp. 361-365, 1975.
	
	
\end{thebibliography}
\end{document}